\documentstyle[preprint,aps,epsf]{revtex}
\begin{document}
\draft
\preprint{
%\vbox{\hbox{ASITP-99-}
%      \hbox{KIAS-P990}
%      \hbox{hep-ph/9910?}
%}
}
\title{
Topcolor-Assisted Supersymmetry
}
\author{ 
Chun Liu
}
\vspace{0.5cm}
\address{
$^a$Institute of Theoretical Physics, Chinese Academy of Sciences,\\ 
PO Box 2735, Beijing 100080, China\\
$^b$School of Physics, Korea Institute for Advanced Study,\\ 
207-43 Cheongryangri-dong, Dongdaemun-gu, Seoul 130-012, Korea
}
\maketitle
\thispagestyle{empty}
\setcounter{page}{1}
\begin{abstract}
It has been known that the supersymmetric flavor changing neutral 
current problem can be avoided if the squarks take the following mass 
pattern, namely the first two generations with the same chirality are 
degenerate with masses around the weak scale, while the third 
generation is very heavy.  We realize this scenario through the 
supersymmetric extension of a topcolor model with gauge mediated 
supersymmetry breaking.\\


\end{abstract}

\newpage

%%%%%%%%%%%%
\section{Introduction}
\label{sec:introduction}
%%%%%%%%%%%%

Supersymmetry (SUSY) \cite{1} provides a solution to the gauge hierarchy 
problem if it breaks dynamically \cite{2}.  However, the general 
supersymmetric extension of the Standard Model (SM) suffers from the 
flavor changing neutral current (FCNC) problem \cite{3}.  The SUSY FCNC 
should be suppressed by certain specific mass patterns of the sleptons 
and squarks.  The sfermion mass pattern depends on the underlying 
physics of SUSY breaking.  For example, one of the popular choice for the 
sfermion mass matrix is the universality in the flavor space.  Such a 
mass matrix can be resulted from the gauge mediated SUSY breaking (GMSB) 
scenario \cite{4,5}.  Here gauge means the SM gauge interactions.  Another
inspiring choice is that the first two generation sfermions are very heavy 
around $(10-100)$ TeV where the third generation sfermions are at the weak
scale \cite{6}.  This kind of model is often referred to as effective SUSY.  
It can be realized in the GMSB scenario \cite{7}, or in that where an 
anomalous U(1) mediates SUSY breaking \cite{8}.  One point in this case 
is that the first two generations and the third generation are treated 
differently.  For example, they maybe in different representations of some
new gauge interactions mediating SUSY breaking.  

In this work, we consider a sfermion mass pattern which looks opposite to 
that of the effective SUSY.  It is that the first two generations with the
same chirality are degenerate with masses around the weak scale, and the 
third generation is super heavy.  The SUSY FCNC is also suppressed in this
case \cite{3}.\footnote{There are other viable alternatives.  For example, 
only the squarks satisfy this mass pattern, while the slepton mass 
matrices follow universality.}  In fact, this pattern is not new.  It 
could be understood by an U(2) symmetry between the first two generations 
in the supergravity scenario \cite{9}.  In this paper an alternative 
origin of it will be discussed.  We note that the above mass pattern can 
be also a result of a supersymmetric topcolor model with GMSB.  Here 
gauge (G) means the gauge interactions of the topcolor model.  

Topcolor models \cite{10} were proposed for a dynamical understanding of 
the third generation quark masses.  The basic idea is that the third 
generation (at least the top quark) undergoes a super strong interaction 
which results in a top quark condensation.  The condensation gives top 
quark a large mass, and the bottom quark mainly gets its mass due to the 
instanton effect of the topcolor interactions.  This top quark 
condensation contributes only a small part of the electroweak symmetry 
breaking (EWSB).  The Higgs mechanism may be introduced for the EWSB.  
Therefore, the idea of the topcolor can be generalized into SUSY models 
naturally.  In the scenario of the GMSB, suppose the strong topcolor gauge 
interaction involves the full third generation, and the first two 
generations participate a weaker gauge interaction universally, the above 
described sfermion mass pattern will be generated.

Note that the decoupling of the third generation scalars is a consistent 
choice in the supersymmetric topcolor models.  Because the third 
generation quarks obtain dynamical masses, the Yukawa couplings are always 
small.  

The whole physical picture is like as follows.  At the energy scale about 
$10^6$ GeV, SUSY breaking occurs in a secluded sector.  It is mediated to 
the observable sector through the gauge interactions.  The scale of the 
messengers is around $10^7$ GeV.  The topcolor scale is around $(1-10)$ 
TeV.  Below this scale, the gauge symmetries break into that of the SM.  
The resultant sparticle spectrum of the observable sector is the following.  
Besides the squarks, the gauginos of the super strong interaction are around 
$100$ TeV.  The gauginos of the weaker gauge interaction are at about the 
weak scale.  The Higgs bosons for the topcolor symmetry breaking are as 
heavy as $100$ TeV, and the Higgs bosons for the EWSB at the weak scale.  
By integrating out the heavy fields above $1$ TeV or so, the 
effective theory is the ordinary (two-Higgs-doublets) topcolor model with 
the weak scale gauginos, Higgsinos and the first two generation squarks 
with degeneracy.

This paper is organized as follows.  After a brief review of the topcolor 
model in the next section.  The supersymmetric extension of the topcolor 
model within the framework of the GMSB is described in Sec. III.  Summary 
and discussions are presented in Sec. IV.

%%%%%%%%%%%
\section{Brief Review of the Topcolor Model}
\label{sec:review}
%%%%%%%%%%%

In this paper, we consider the topcolor model which, at the scale about 
$(1-10)$ TeV, has interactions \cite{10} 
SU(3)$_1\times$SU(3)$_2\times$U(1)$_{Y_1}\times$U(1)$_{Y_2}\times$SU(2)$_L$.
The fermions are assigned
(SU(3)$_1$, SU(3)$_2$, U(1)$_{Y_1}$, U(1)$_{Y_2}$) quantum numbers as 
follows, 
\begin{eqnarray}
\label{1}
(t, b)_L&\sim& (3\;, 1\;, \frac{1}{3} \;, 0) \; ,~~~
(t, b)_R\sim (3\;, 1\;, (\frac{4}{3} \;, -\frac{2}{3}) \;, 0)\;,
\nonumber\\
(\nu_{\tau}\;, \tau)_L&\sim& (1\;, 1\;, -1\;, 0) \; ,~~~
\tau_R\sim (1\;, 1\;, -2\;, 0)\;,\nonumber\\
(u\;, d)_L\;, (c\;, s)_L&\sim& (1\;, 3\;, 0\;, \frac{1}{3})\;,~~~
(u\;, d)_R\;, (c\;, s)_R\sim (1\;, 3\;, 0\;, (\frac{4}{3}\;, 
-\frac{2}{3}))\;, \nonumber\\
(\nu\;, l)_L(l=e, \mu)&\sim& (1\;, 1\;, 0\;, -1) \; ,~~~
l_R\sim (1\;, 1\;, 0\;, -2)\; .
\end{eqnarray}
The topcolor symmetry breaks spontaneously to SU(3)$_1\times$SU(3)$_2\to$
SU(3)$_{\rm QCD}$ and U(1)$_{Y_1}\times$U(1)$_{Y_2}\to$U(1)$_Y$ through an
scalar field $\phi(3\;, \bar{3}\;, \frac{1}{3}\;, -\frac{1}{3})$ which 
develops a vacuum expectation value (VEV).  The SU(3)$_1\times$U(1)$_{Y_1}$ 
are assumed to be strong which make the formation of a top quark condensate 
but disallow the bottom quark condensate.  The bottom quark mainly gets its 
mass due to the SU(3)$_1$ instanton effect.  The $\tau$ lepton does not 
condensate.

%%%%%%%%%%%
\section{Supersymmetric Topcolor Model}
\label{sec:susyt}
%%%%%%%%%%%

In the supersymmetric extension, the gauge symmetries of the above topcolor 
model keep unchanged.  The particle contents are given below.  In addition 
to the superpartners of the particles described in the last section, some 
elementary Higgs superfields are introduced.  The breaking of the topcolor 
symmetry needs one pair of the Higgs superfields $\Phi_1$ and $\Phi_2$.  
And the EWSB requires another pair of the Higgs superfields $H_u$ and 
$H_d$, like in the ordinary supersymmetric SM.  Their quantum numbers under 
the 
SU(3)$_1\times$SU(3)$_2\times$U(1)$_{Y_1}\times$U(1)$_{Y_2}\times$SU(2)$_L$
are
\begin{eqnarray}
\label{3}
\Phi_1(3\;, \bar{3}\;, \frac{1}{3}\;, -\frac{1}{3}\;, 0)\;, &~~~&
\Phi_2(\bar{3}\;, 3\;, -\frac{1}{3}\;, \frac{1}{3}\;, 0)\;; \nonumber\\
H_u(1\;, 1\;, 0\;, 1\;, 2)\;, &~~~& H_d(1\;, 1\;, 0\;, -1\;, 2)\;.
\end{eqnarray}
The messenger sector is introduced as 
\begin{eqnarray}
\label{3}
S_1\;, S'_1 &=& (1\;, 1\;, 1\;, 0\;, 2)\;, ~~~
\bar{S_1}\;, \bar{S'_1} = (1\;, 1\;, -1\;, 0\;, 2)\;, \nonumber\\
T_1\;, T'_1 &=& (3\;, 1\;, -\frac{2}{3}\;, 0\;, 1)\;, ~~~
\bar{T_1}\;, \bar{T'_1} = (\bar{3}\;, 1\;, \frac{2}{3}\;, 0\;, 1)\;, 
\end{eqnarray}
and
\begin{eqnarray}
\label{4}
S_2\;, S'_2 &=& (1\;, 1\;, 0\;, 1\;, 2)\;, ~~~
\bar{S_2}\;, \bar{S'_2} = (1\;, 1\;, 0\;, -1\;, 2)\;, \nonumber\\
T_2\;, T'_2 &=& (1\;, 3\;, 0\;, -\frac{2}{3}\;, 1)\;, ~~~
\bar{T_2}\;, \bar{T'_2} = (1\;, \bar{3}\;, 0\;, \frac{2}{3}\;, 1)\;.  
\end{eqnarray}
Compared to Ref. \cite{4}, we have introduced an extra set of messengers 
so as to mediate the SUSY breaking to both SU(3)$_1\times$U(1)$_{Y_1}$ and
SU(3)$_2\times$U(1)$_{Y_2}$.  Furthermore, there are three gauge-singlet
superfields, $X$, $Y$ and $Z$.  $Y$ is responsible for the SUSY breaking, 
$X$ is related to the EWSB, and $Z$ to the topcolor symmetry breaking.  

The superpotential is written as follows,
\begin{eqnarray}
\label{5}
{\cal W}&=&m_1(\bar{S'_1}S_1+S'_1\bar{S_1})+m_2(\bar{T'_1}T_1
+T'_1\bar{T_1})+m_3S_1\bar{S_1}+m_4T_1\bar{T_1} \nonumber\\ 
&&m_1'(\bar{S'_2}S_2+S'_2\bar{S_2})+m_2'(\bar{T'_2}T_2
+T'_2\bar{T_2})+m_3'S_2\bar{S_2}+m_4'T_2\bar{T_2} \nonumber\\
&&+Y(\lambda_1S_1\bar{S_1}+\lambda_2T_1\bar{T_1}
+\lambda_1'S_2\bar{S_2}+\lambda_2'T_2\bar{T_2}-\mu_1^2) \nonumber\\
&&+\lambda_3X(H_uH_d-\mu_2^2)+\lambda_4Z[{\rm Tr}(\Phi_1\Phi_2)-\mu_3^2]
\; ,
\end{eqnarray}
where the Yukawa interactions are not written.  It is required that 
$m_3^{(\prime)}/m_4^{(\prime)}\not=\lambda_1^{(\prime)}/
\lambda_2^{(\prime)}$ so that the terms proportional to $m_3^{(\prime)}$ 
and $m_4^{(\prime)}$ cannot be eliminated by a shift in $Y$.  

The model conserves the number of $S_i$-type and $T_i$-type ($i=1\;, 2$) 
fields.  In addition, the superpotential has a discrete symmetry of 
$(\bar{S_i}^{(\prime)}\;, \bar{T_i}^{(\prime)})
\leftrightarrow(S_i^{(\prime)}\;, T_i^{(\prime)})$.
The way of introducing the singlet fields $X$, $Y$ and $Z$ more naturally 
was discussed in Ref. \cite{11} where these kind of fields are taken to be 
composite.  Moreover, the Fayet-Iliopoulos D-terms for the U(1) charges 
have been omitted.  This is natural in the GMSB scenario.  
The above discrete symmetry and the exchange symmetry of $\Phi_1$ and 
$\Phi_2$ in the superpotential avoid such D-terms at the one-loop order.

The SUSY breaking is characterized by the term $\mu_1^2Y$ in Eq. (\ref{5}).  
It is communicated to the observable sector through the gauge interactions 
by the messengers.  The SU(3)$_2\times$U(1)$_{Y_2}\times$SU(2)$_L$ are 
weak enough to be described in perturbation theory.  Their gauginos acquire 
masses in the one-loop order \cite{4,12},
\begin{eqnarray}
\label{6}
M_{\lambda_{\rm SU(3)_2}}&=&\frac{\alpha_3'}{4\pi}{\cal M}_T \; ,
\nonumber\\[3mm]
M_{\lambda_{\rm U(1)_{Y_2}}}&=&\frac{\alpha_1'}{4\pi}
({\cal M}_S+\frac{2}{3}{\cal M}_T) \; , \nonumber\\[3mm]
M_{\tilde{W}}&=&\frac{\alpha_2}{4\pi}{\cal M}_S \; ,
\end{eqnarray}
where $\alpha_i^{(\prime)}=g_i^{(\prime)2}/4\pi$ with $g_3'$, $g_1'$ and $g_2$ 
being the gauge coupling constants of the 
SU(3)$_2\times$U(1)$_{Y_2}\times$SU(2)$_L$.  For simplicity, we take 
$m_1^2\sim m_2^2\sim m_1^{\prime 2}\sim m_2^{\prime 2}\gg
\lambda_{1, 2}^{(\prime)}\mu_1^2\gg m_3^{(\prime)2}\sim m_4^{(\prime)2}$ and 
$\lambda_1\sim\lambda_2\sim\lambda_1'\sim\lambda_2'\sim O(1)$.  In this case, 
the ${\cal M}_S$ and ${\cal M}_T$ are approximately $\lambda_1\mu_1^2/m_1$.  
They are about $100$ TeV by choosing, say 
$\mu_1\sim 10^6$ GeV and $m_1\sim 10\mu_1$.  For the 
SU(3)$_1\times$U(1)$_{Y_1}$, however, the gaugino masses cannot be 
calculated by the perturbation method, because the interactions are too 
strong.  Nevertheless they should be at the order of 
$\lambda_1\mu_1^2/m_1$ given the above parameter choice,
\begin{equation}
\label{7}
M_{\lambda_{\rm SU(3)_1}}\sim M_{\lambda_{\rm U(1)_{Y_1}}}\sim 100 
~~{\rm TeV} \; .
\end{equation}
Similarly, the first two generation scalar quarks and the electroweak 
Higgs particles obtain their masses in the two-loop order, 
\begin{eqnarray}
\label{8}
m_{\tilde{Q}_1}^2&=&m_{\tilde{Q}_2}^2\simeq
\frac{4}{3}\left(\frac{\alpha_3'}{4\pi}\right)^2\Lambda_T^2
+\frac{3}{4}\left(\frac{\alpha_2}{4\pi}\right)^2\Lambda_S^2
+\frac{1}{4}\left(\frac{\alpha_1'}{4\pi}\right)^2(\Lambda_S^2
+\frac{2}{3}\Lambda_T^2) \; , \nonumber \\
m_{\tilde{c}_R}^2&=&m_{\tilde{u}_R}^2\simeq
\frac{4}{3}\left(\frac{\alpha_3'}{4\pi}\right)^2\Lambda_T^2
+\frac{4}{9}\left(\frac{\alpha_1'}{4\pi}\right)^2(\Lambda_S^2
+\frac{2}{3}\Lambda_T^2) \; , \nonumber \\
m_{\tilde{s}_R}^2&=&m_{\tilde{d}_R}^2\simeq
\frac{4}{3}\left(\frac{\alpha_3'}{4\pi}\right)^2\Lambda_T^2
+\frac{1}{9}\left(\frac{\alpha_1'}{4\pi}\right)^2(\Lambda_S^2
+\frac{2}{3}\Lambda_T^2) \; , \nonumber \\
m_{h_u}^2&=&m_{h_d}^2\simeq
\frac{3}{4}\left(\frac{\alpha_2}{4\pi}\right)^2\Lambda_S^2
+\frac{1}{4}\left(\frac{\alpha_1'}{4\pi}\right)^2(\Lambda_S^2
+\frac{2}{3}\Lambda_T^2) \; , 
\end{eqnarray}
where $Q_1$ and $Q_2$ stand for the superfields of $(u\;, d)_L$ and 
$(c\;, s)_L$ respectively.  And $(h_u\; h_d)$ are the scalar 
components of $(H_u\;, H_d)$.  $\Lambda_S^2$ and $\Lambda_T^2$ was 
calculated to be \cite{4}
\begin{equation}
\label{9}
\Lambda_S^2=\frac{4\lambda_1'^2\mu_1^4}{m_1'^2} \; , ~~~
\Lambda_T^2=\frac{4\lambda_2'^2\mu_1^4}{m_2'^2} \; .
\end{equation}
For the third generation squarks and the topcolor Higgs' $\phi_1$ and 
$\phi_2$, the masses are       
around $\Lambda_S^2$ or $\Lambda_T^2$, 
\begin{equation}
\label{10}
m_{\tilde{Q}_3}^2\sim m_{\tilde{t}_R}^2\sim m_{\tilde{b}_R}^2\sim 
m_{\phi_1}^2=m_{\phi_2}^2\sim \Lambda_S^2 \; ,\Lambda_T^2\sim 
(100 ~~{\rm TeV})^2 \; .
\end{equation}

We have seen that for the super strong topcolor interactions, the 
relevant supersymmetric particles are super heavy $\sim 100$ TeV so that 
they decouple at the topcolor scale.  The topcolor physics does not 
change even after the supersymmetric extension.  However the topcolor 
Higgs fields seem to be too heavy.  

Let us consider the breaking of the gauge symmetries.  The 
SU(3)$_1\times$SU(3)$_2\times$U(1)$_{Y_1}\times$U(1)$_{Y_2}$ break into 
the diagonal subgroups SU(3)$_{\rm QCD}\times$U(1)$_Y$ when the Higgs 
fields $\phi_1$ and $\phi_2$ get non-vanishing VEVs,
\begin{equation}
\label{11}
\langle\phi_1\rangle=v_1\left(\begin{array}{ccc}
1 &0 &0 \\
0 &1 &0\\
0 &0 &1
\end{array}
\right)
~~~{\rm and}~~~ 
\langle\phi_2\rangle=v_2\left(\begin{array}{ccc}
1 &0 &0 \\
0 &1 &0\\
0 &0 &1
\end{array}
\right) \; .
\end{equation}
$v_1$ and $v_2$ are determined by the minimum of the following potential,
\begin{equation}
\label{12}
V_{topc}=|\lambda_4(3v_1v_2-\mu_3^2)|^2+\frac{g_1^2+g_1^{\prime 2}}{2}
(v_1^2-v_2^2)^2+m_{\phi_1}^2v_1^2+m_{\phi_2}^2v_2^2 \; ,
\end{equation}
where $g_1$ is the coupling constant of the U(1)$_{Y_1}$.  It is easy to 
see that in the case of $\lambda_4\mu_3^2\geq m_{\phi_i}^2$, 
\begin{equation}
\label{13}
v_1=v_2=\frac{1}{\sqrt{3}}\left(\mu_3^2-\frac{m_{\phi_i}^2}{\lambda_4}
\right)^{1/2} \; .
\end{equation}
To keep $v_1$ and $v_2$ to be around a few TeV, certain fine-tuning for 
the scale $\mu_3$ is required in this model to cancel the $100$ TeV 
$m_{\phi_i}$, where the coupling $\lambda_4$ is $O(1)$.  The value of 
$\mu_3$ is more natural if the topcolor 
scale is higher, such as $10$ TeV.  
However, it should be noted that raising topcolor scale makes the 
effective topcolor theory more tuned.  

At the energy below the topcolor scale, the model is described by an 
effective theory in which the gauge symmetry groups are that of the SM, 
and there are two Higgs doublets and three generation quarks with 
four-fermion topcolor interaction for the third generation.  In addition,
there are weak scale gauginos of the SM, squarks of the first and second 
generations and doublet Higgsinos which become massive after the EWSB.  
There are also topcolor Higgsinos of $\Phi_1$ and $\Phi_2$ after the 
topcolor symmetry breaking.  They typically have $(1-10)$ TeV mass and 
are not expected to be important to the low energy physics.  Because of 
the degeneracy between the first two generation squarks and the 
decoupling of the third generation squarks, this model is free from the 
SUSY FCNC problem.  

The physics of the topcolor four-fermion interaction and the EWSB in 
this model is essentially the same as that without SUSY, which will not be 
discussed further.

%%%%%%%%%%%
\section{Summary and Discussion}
\label{sec:summary}
%%%%%%%%%%%

It has been known that the SUSY FCNC problem can be avoided if the squarks 
take the mass pattern that the first two generations with the same 
chirality are degenerate and the third generation is super heavy.  We have 
constructed a supersymmetric topcolor model within GMSB to realize this 
mass pattern.  The pattern is stable under the correction of the Yukawa 
interactions because they are weak and the third generation quarks obtain 
masses dynamically.

This model has therefore, the phenomenologies of both SUSY and topcolor.  
It predicts weak scale SUSY particles, like the SM gauginos, Higgsinos.
It also predicts top pions.  These predictions can be tested directly in 
the experiments in the near future.
The indirect evidences of this model in low energy processes, such as in 
the B decays \cite{13}, and the $R_b$ problem of it \cite{14} are more 
complicated because of the involvement of 
both the SUSY and the topcolor, and deserve a separate study.  

It should be addressed that this model has an inherent tuning problem.  
This required tuning follows from the very large masses ($\sim 100$ TeV) 
of the third generation scalars and the topcolor Higgs.  These fields are 
closely related to the topcolor and the EWSB scales which are, however, 
lower than 100 TeV.  We have explicitly mentioned the tuning below 
Eq. (\ref{13}).  Another aspect of this tuning is that the naturalness of 
the EWSB requires the third generation scalars to be lighter than 20 TeV 
\cite{6}.  Note that the large mass 100 TeV is just a rough estimate due 
to that we are lack of nonperturbative calculation method.  On the other 
hand, if we adjust the SUSY breaking scale and the messenger scale to be 
somewhat lower than what we have chosen, this problem can be less severe.  

We emphasis that it is the degeneracy of the first two generations, rather 
than the heaviness of the third generation, that plays the essential role 
in solving the SUSY FCNC problem.  In this sense, the consideration of this 
paper is less nontrival than the idea of effective SUSY.  However, if we 
further consider the underlying theory, the models which realize effective 
SUSY \cite{7,8} and the SUSY topcolor model of this paper are on an equal 
footing.  

A comment should be made on the necessity of the supersymmetric 
topcolor.  Although SUSY does not necessarily need the help from topcolor, 
their combination has certain advantages.  As is well-known, SUSY keeps 
the weak scale, but cannot explain it.  The weak scale may have a 
dynamical origin \cite{15,16,11,17}.  In this case, it is natural to 
expect that the physics which explains the fermion masses is also at some 
low energy.  Topcolor provides such physics for the hierarchy between the 
third generation and the first two generations.  On the other hand, SUSY 
maybe helpful to understand the hierarchy between the first and the second 
generation further.  For instance, it is possible that the second 
generation quarks mainly get their masses from the electroweak Higgs VEVs, 
and the first generation quarks purely from the sneutrino VEVs \cite{18}.  

Finally, it should be noted that the very heavy third generation squarks 
may pull up the light scalars.  This pull up occurs through two- or 
more-loop diagrams with the topcolor Higgs exchanges.  The heavy topcolor 
Higgs suppress this quantum correction.  The suppression, however, may be 
not enough to keep the results of Eq. (\ref{8}) from significant changing 
numerically.  The fine-tuning problem which was discussed above re-appears 
here.  In fact, the drawback of the SUSY and topcolor combination is that 
the SUSY breaking scale and the topcolor scale are irrelevant.  It might be 
hopeful to think of certain dynamics to make relation between them.  For 
example, it is possible that the topcolor Higgs superfields are also the 
SUSY breaking messengers.  This possibility will simplify the model and 
reduce the fine-tuning.
It is reasonable to say that the supersymmetric topcolor is an interesting 
scenario which is worthy to be studied further.

%%%%%%%%%%%%%%%%%%%%%%%%%%%%%%%%%%%%%%%%%%%%%%%%
\acknowledgments
%%%%%%%%%%%%%%%%%%%%%%%%%%%

The author would like to thank many Korean colleagues for various helpful 
discussions.

%\newpage


\begin{thebibliography}{99}

\bibitem{1}
J. Wess and B. Zumino, Nucl. Phys. {\bf B70}, 39 (1974);\\
Y. Gol'fand and E. Likhtam, JETP Lett. {\bf 13}, 323 (1971);\\
D.V. Volkov and V. Akulov, Phys. Lett. {\bf B46}, 109 (1973).

\bibitem{2}
E. Witten, Nucl. Phys. {\bf B188}, 513 (1981).

\bibitem{3}
For example, see F. Gabbiani, E. Gabrielli, A. Masiero and L. Silvestrini, 
Nucl. Phys. {\bf B477}, 321 (1996).

\bibitem{4}
M. Dine and W. Fischler, Phys. Lett. {\bf B110}, 227 (1982);\\
L. Alvarez-Gaum\'e, M. Claudson and M. Wise, Nucl. Phys. {\bf B207}, 96 
(1982);\\ 
C.R. Nappi and B.A. Ovrut, Phys. Lett. {\bf B113}, 175 (1982).

\bibitem{5}
M. Dine and A.E. Nelson, Phys. Rev. {\bf D48}, 1277 (1993);\\
M. Dine, A.E. Nelson and Y. Shirman, Phys. Rev. {\bf D51}, 1362 (1995);\\
M. Dine, A.E. Nelson, Y. Nir and Y. Shirman, Phys. Rev. {\bf D53}, 2658 
(1996).

\bibitem{6}
A.G. Cohen, D.B. Kaplan and A.E. Nelson, Phys. Lett. {\bf B388}, 588 
(1996);\\
S. Dimopoulos and G.F. Giudice,  Phys. Lett. {\bf B357}, 573 (1995).

\bibitem{7}
S. Ambrosanio and A.E. Nelson, Phys. Lett. {\bf B411}, 283 (1997).

\bibitem{8}
G. Dvali and A. Pomarol, Phys. Rev. Lett. {\bf 77}, 3728 (1996).

\bibitem{9}
A. Pomarol and D. Tommasini, Nucl. Phys. {\bf B466}, 3 (1996);\\
R. Barbieri, G. Dvali and L.J. Hall, Phys. Lett. {\bf B377}, 76 (1996).

\bibitem{10}
C.T. Hill, Phys. Lett. {\bf B266}, 419 (1991);
Phys. Lett. {\bf B345}, 483 (1995);\\
For a review, see C.T. Hill, hep-ph/9702320.

\bibitem{11}
C. Liu, hep-ph/9903395.

\bibitem{12}
For reviews, see G.F. Giudice and R. Rattazzi, hep-ph/9801271;\\
S.L. Dubovsky, D.S. Gorbunov and S.V. Troisky, hep-ph/9905466.

\bibitem{13}
G. Buchalla, G. Burdman, C.T. Hill and D. Kominis, Phys. Rev. {\bf D53}, 
5185 (1996);\\
K.Y. Lee, Phys. Rev. {\bf D57}, 598 (1998);\\
Z. Xiao {\it et al.}, Eur. Phys. J. {\bf C10}, 51 (1999); hep-ph/9903346.

\bibitem{14}
G. Burdman and D. Kominis, Phys. Lett. {\bf B403}, 101 (1997);\\
C.T. Hill and X. Zhang, Phys. Rev. {\bf D51}, 3563 (1995);\\
W. Loinaz and T. Takeuchi, Phys. Rev. {\bf D60}, 015005 (1999).

\bibitem{15}
S. Weinberg, Phys. Rev. {\bf D19}, 1277 (1978);\\
L. Susskind, Phys. Rev. {\bf D20}, 2619 (1979).

\bibitem{16}
M. Dine, W. Fischler and M. Srednicki, Nucl. Phys. {\bf B189}, 575 
(1981);\\
S. Dimopoulos and S. Raby, Nucl. Phys. {\bf B192}, 353 (1981).

\bibitem{17}
K. Choi and H.D. Kim, hep-ph/9906363.

\bibitem{18}
C. Liu, Int. J. Mod. Phys. {\bf A11}, 4307 (1996);\\
C. Liu and J. Song, Phys. Rev. {\bf D60}, 036002 (1999);\\
J.E. Kim, B. Kyae and J.S. Lee, Phys. Lett. {\bf B447}, 110 (1999).

\end{thebibliography}
\end{document}